\documentclass{aa}
\usepackage{graphicx}

\input{psfig}

\newcommand{\HI}{H{\,\small I}}

\newcommand{\ie}{{\sl i.e.}}

\newcommand{\etal}{{\sl et al.\ }}

\newcommand{\kms}{km s$^{-1}$}
\newcommand{\whz}{W Hz$^{-1}$}

\newcommand{\ltae}{\raisebox{-0.6ex}{$\,\stackrel
{\raisebox{-.2ex}{$\textstyle <$}}{\sim}\,$}}
\newcommand{\gtae}{\raisebox{-0.6ex}{$\,\stackrel
{\raisebox{-.2ex}{$\textstyle >$}}{\sim}\,$}}

\begin{document}
%\thesaurus{ 06() }

\title{A sample of southern Compact Steep Spectrum radio sources: \\
The VLBI observations \thanks{Based on observations with the Southern 
Hemisphere VLBI Network (SHEVE) and the MERLIN.  
}}

\authorrunning{ Tzioumis et al.}
\titlerunning{\small VLBI observations of southern CSS radio sources}

\subtitle{}

\author{
A.  Tzioumis\inst{1}, E.  King\inst{6,11}, R.  Morganti\inst{3}, D. 
Dallacasa\inst{2,9}, C.  Tadhunter\inst{4}, C.  Fanti\inst{2,10}, J. 
Reynolds\inst{1}, D.  Jauncey\inst{1}, R. Preston\inst{5}, P. 
McCulloch\inst{6}, S.  Tingay\inst{1}, P. Edwards\inst{7,12}, M.  Costa\inst{6},
D.  Jones\inst{5}, J.  Lovell\inst{6,1}, R.  Clay\inst{7}, D.  Meier\inst{5}, D. 
Murphy\inst{5}, R.  Gough\inst{1}, R.  Ferris\inst{1}, G.  White\inst{8}, P. 
Jones\inst{8}}

\institute {ATNF -- CSIRO, Epping, Australia \and
 Istituto di Radioastronomia -- CNR, via Gobetti 101, I-40129 Bologna,
 Italy \and 
Netherlands Foundation for Research in Astronomy, Postbus 2, NL-7990 AA,
Dwingeloo, The Netherlands
\and
 Department of Physics, University of Sheffield, UK \and
 Jet Propulsion Laboratory, Caltech, Pasadena, CA, USA \and
 University of Tasmania, Hobart, Tasmania, Australia \and
 University of Adelaide, Adelaide, SA, Australia \and
 University of Western Sydney, Kingswood, NSW, Australia \and
 Astronomy Dept. Universit\`a di Bologna, via Ranzani 1, I-40127
 Bologna, Italy \and 
 Physics Dept. Universit\`a di Bologna, via Irnerio 46, I-40126
 Bologna, Italy \and
COSSA/EOC -- CSIRO, Canberra, Australia \and
ISAS, Sagamihara, Kanagawa, Japan
}

\offprints{A.Tzioumis}

\date{Received 11-03-02; accepted 26-06-02}

\markboth{VLBI Observations of CSS Radio Sources}{Tzioumis  et al.}

\abstract{ A small sample of 7 southern Compact Steep Spectrum (CSS)
radio sources has been selected as part of the study of a larger flux-limited
complete sample of radio sources. High resolution images, using the VLBI
network in the southern hemisphere and the high resolution
MERLIN array, are presented for all
sources in the CSS sample. The overall morphology of each source consists of
well-defined double lobes but with substantial diffuse and extended components
present. In the majority of cases only a fraction of the total flux density is
detected on the VLBI baselines, indicating the presence of larger extended
radio structures. However, all sources are unresolved at arcsecond scales and
are of sub-galactic size, with linear size in the range 0.1--2~kpc. The radio
properties of the sources agree well with CSS sources in other samples.
}

\maketitle

\section{Introduction}

Among the many open questions about powerful radiogalaxies is the
early evolution of these objects.  This is an important issue not only
from the point of view of the detailed phenomenology of these objects
but also for understanding the evolution of massive galaxies.
Galaxies in their early stage of radio activity are likely to have
their nuclear regions enshrouded in a cocoon or thick disk of material
left over from the event(s) that trigger the activity. The radio jet
expanding in this medium will interact and sweep out this rich
interstellar medium (ISM). These processes have been suggested to
profoundly affect even the star formation history in luminous galaxies
(Silk \& Rees 1998). Thus, the study of the early-phase of radio
activity and its effect on the galaxy medium has broad implications.
 
Compact Steep Spectrum (CSS) and Gigahertz Peaked Spectrum (GPS) radio
sources have now been recognised to be the likely candidates of radio
sources in such an early stage of their life. These radio sources are
powerful (P$_{2.7 {\rm GHz}} \gtae 10^{25.5}$ W~Hz$^{-1}$), have a
steep spectral index at high frequencies ($\alpha < -0.5$, $S \sim
\nu^{\alpha}$) and sub-galactic size ($<$20 kpc). Indeed, the
turnover often observed in their spectra (around 1 GHz for the GPSs and
about 100 MHz for the CSSs) is usually interpreted as synchrotron
self-absorption, with the frequency of this turnover varying with the
linear scale of the source, although free-free absorption may also
play a role (Bicknell et al. 1997). A review of the characteristics of
CSS/GPS sources at different wavelengths is given in O'Dea (1998).

The age of CSS/GPS  sources has been recently estimated using both the
lobe proper motion (e.g. Owsianik \& Conway 1998) and the radiative
ages (Murgia et al. 1999). The ages derived are less than a few
thousand  years, therefore strongly supporting the {\sl ``youth''
scenario} in the interpretation of these sources. A summary of the
results  is given in the review by Fanti et al. (2000).  

The sub-galactic size and the young age of these sources makes them an
extremely interesting class of radio source. They can be used to
probe further the characteristics of the ISM in the early phase of
radio source evolution and the effects of the radio plasma expanding in 
this medium. This has been done in a restricted number of objects both
using the atomic hydrogen (e.g. Conway 1996, Pilstr\"om 2001, Morganti
et al. 2001, Peck \& Taylor 2001) and the ionised gas (e.g. Gelderman \&
Whittle 1994, Morganti
et al. 1997, Axon et al. 2000, O'Dea et al. 2002).

An other important aspect in the study of these objects is the
comparison of their characteristics with those of the extended radio
galaxies.  This has been done quite extensively in the radio to
investigate the evolutionary scenario. At wavelengths different from
radio the available data are still limited. ISO observations (Fanti et
al. 2000) and optical and near IR data (de Vries et al. 1998) showed
no significant difference between CSS/GPS sources and extended radio
galaxies with similar radio power, failing therefore to detect any
extra component capable to ``frustrate'' the CSS/GPS sources. 

To extend the comparison between the optical properties of CSS/GPS and
extended radio sources, a group of CSS/GPS sources was selected from a
complete sub-sample of the {\sl 2~Jy sample of radio sources} (see Sect.
2 for details). Optical spectra were obtained for both the compact as
well as for the extended sources in this sample (see Tadhunter et
al. 1993) and the spectral characteristics could therefore be compared
for objects in a similar range of redshift and radio power.  A
first-order similarity was found between the spectra of CSS/GPS
sources and the extended source of similar radio power although the
former may be of slightly lower ionisation, possible evidence for the
compression effect of shocks due to jet/cloud interaction (Morganti et
al. 1997).  For the 2~Jy
CSS/GPS source sample however, there is a lack of high resolution radio
images capable of providing information on their morphology. Most of the
well studied CSS/GPS galaxies tend to have symmetric, double-lobed
radio structures (e.g. Dallacasa et al. 1995, O'Dea 1998), reminiscent
of scale-down versions of extended, edge-brightened Fanaroff-Riley II
sources.  Collecting this information for the CSS/GPS sources in the
2~Jy sample is particularly important also for the interpretation of
their optical data. The presence of more complex radio morphology in
these sources (or in some of them) would be an indication of a
particularly strong interaction between the radio plasma and the ISM,
more than usually found in this type of radio source.

In order to fill this gap, in this paper we present and discuss new high
resolution (VLBI and MERLIN) data for this group of seven 2~Jy CSS/GPS
sources. These radio images permit a proper morphological
classification of the sub-kpc radio emission in these objects.

\begin{figure}
\centerline{\psfig{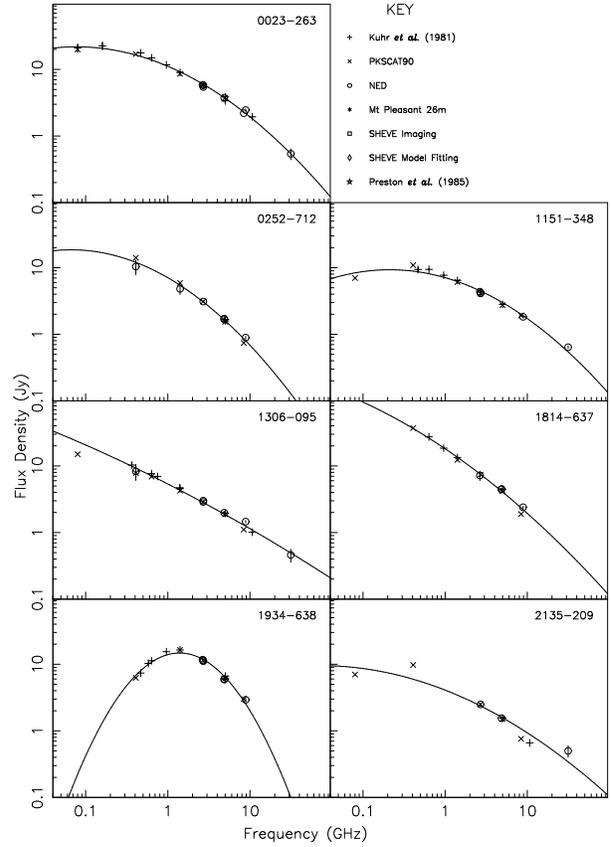}}
\caption{Radio spectral index plots for all sources. \label{fig:spectra}}
\end{figure}

\begin{table}
\begin{center}
\begin{tabular}{lcccccc}
\hline\hline
Name  & ID & $m_V$ & $z$ &  scale  & $\log P_{2.7~GHz}$ & $\alpha^{8.4}_{2.7}$ \\ 
      &    & &   &  pc/mas    &  \whz\  & \\
\hline  
           &   &      &       &     &       &      \\
$0023-263$ & G & 19.5 & 0.322 & 2.9 & 26.91 & -0.8 \\
$0252-712$ & G & 20.9 & 0.566 & 3.7 & 27.28 & -1.25\\
$1151-348$ & Q & 17.8 & 0.258 & 2.5 & 26.56 & -0.7 \\
$1306-095$ & G & 20.5 & 0.464 & 3.5 & 26.94 & -0.85\\
$1814-637$ & G & 18.0 & 0.063 & 0.8 & 25.54 & -1.2 \\
$1934-638^*$ & G & 18.4 & 0.183 & 2.0 & 26.68  & -1.15\\
$2135-209$ & G & 19.4 & 0.635 & 3.9  & 27.20 & -1.05 \\
\hline
\multicolumn{6}{l}{\small $^*$ Gigahertz-Peaked Spectrum source (GPS)}
\end{tabular}
\end{center}
\caption{CSS sources in the sample  with a summary of their
characteristics.} 
\label{tab:srclist}
\end{table}

\begin{figure}
\centerline{\psfig{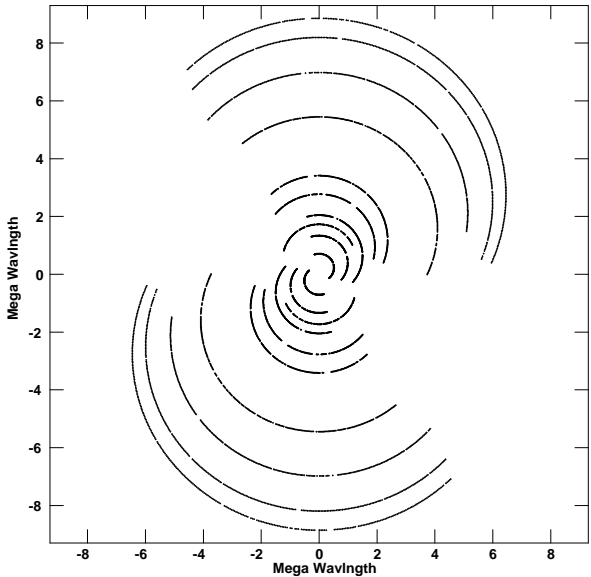}}
\caption{Typical uv coverage of the SHEVE array, from the 0252-712 at 2.3 
GHz. (No detections to Perth or Hartebeesthoek).     \label{fig:uv}}
\end{figure}

\section{The Sample}

The group of CSS/GPS sources considered here has been selected from
the complete sub-sample of the Wall \& Peacock 1985 (WP85) 2~Jy sample
described in Tadhunter et al. 1993.  This sub-sample is formed by 88
radio sources (radio galaxies and quasars) having S$_{\rm 2.7GHz}>$2.0 Jy,
$\delta<10^\circ$ and $z<$0.7. The sample has been studied extensively
at various wavelengths: optical (Tadhunter et al. 1993, 1998), radio
(Morganti et al. 1993, 1997, 1999, Venturi et al. 2000) and X-ray
(Siebert et al. 1996, Padovani et al. 1999, Trussoni et al. 1999).

According to the definition in Fanti et al. 1995, a source is
classified as CSS when at high frequency $\alpha<-0.5$
($S \sim \nu^{\alpha}$) and the linear dimensions are
$\le$~~15kpc\footnote{H$_\circ = 100$ \kms\ Mpc$^{-1}$, q$_\circ$=0.5}.
To make the selection consistent with previous works of Fanti et al.
1990 and Spencer et al. 1991, only objects with
P$_{\rm 2.7~GHz}>10^{25.5}$~W~Hz$^{-1}$ were selected.  With these
criteria we find 7 CSS (6 galaxies and 1 quasar) in our complete
sample \ie\ 8\% of the total number of sources or 12.5\% of the
P$_{\rm 2.7~GHz}>10^{25.5}$~W~Hz$^{-1}$ sub-sample.

The seven CSS sources in our sample are listed in
Table~\ref{tab:srclist} together with some additional information.
Their radio power is consistent with the typical radio power for this
class of objects. Optical identifications and redshifts (WP85; di
Serego Alighieri
et al. 1994) are available for every source in this sample.  All
sources with declination north of $-40^\circ$ are unresolved in VLA
images ($\theta < 3^{\prime\prime}$) and the remaining are part of the
ATCA calibration list ($\theta < 1^{\prime\prime}$). Radio spectra
from the present radio observations and from data in the literature
(King 1994) are presented in Fig.~\ref{fig:spectra}. The two-point
spectral index $\alpha^{8.4}_{2.7}$ listed in Table~\ref{tab:srclist}
clearly demonstrates the steep spectrum nature of these objects.

\section{Observations, data reduction and analysis}

\begin{table*}
\begin{center}
\begin{tabular}{clrrr} \hline \hline
Key &Antenna    & \multicolumn{1}{c}{Size} & \multicolumn{2}{c}{SEFD (Jy)} \\
                                        \cline{4-5}
    &                   & (m)          & 2.3 GHz    & 8.4 GHz  \\ \hline
P   & Parkes(ATNF)           & 64           &   90       &  90    \\
M   & Mopra (ATNF)$^*$       & 22           &  400       & 400    \\
C   & Narrabri-CA (ATNF)     & 22           &  400       & 400    \\
3   & Tidbinbilla (DSS43)    & 70           &   15       &  20    \\
5   & Tidbinbilla (DSS45)    & 34           &  165       & 130    \\
2   & Tidbinbilla (DSS42)    & 34           &  100       & 130    \\
H   & Hobart (Univ.Tasmania) & 26           &  750       & 650    \\
G   & Perth (ESA)            & 15           & 3300       &  --    \\
E   & Harteebeesthoek        & 26           &  400       & 950    \\
\hline\hline
\multicolumn{4}{l}{$^*$ \footnotesize The Mopra antenna became
 operational in late 1991.}
\end{tabular}
\end{center}
\caption{Antennas participating in the VLBI observations. Note
         that the three Tidbinbilla antennas are at essentially
         the same location and only one antenna is used at any one time,
         depending on availability.
         The SEFD (system equivalent flux density)
         is given by the ratio between system temperature ($^\circ$K)
         and gain $^\circ$K/Jy.}
\label{tab:ants}
\end{table*}

\begin{table*}
\begin{center}
  \begin{tabular}{cccclccccc} 
\hline \hline
Source   & Epoch   & Freq. & Duration &Stations&$B_{max}$* & Beam &Dynamic&
         $\sigma$ & Figure \\
         &         & GHz &  (hours) &        & $M\lambda$ & mas ~ $^\circ$ 
&Range
         &{\small mJy/beam} &  Ref.   \\ \hline
0023$-$263 & 1991.34 & 2.29  &   11.5   & 32PHC    & 10.5 & ~31$\times$14 
~-82    & 60:1  &4.5        
         & \ref{fig:img0023sx} \\
           & 1991.90 & 8.42  &   11     & 32PHCM   & 29.9 & ~13$\times$5~ 
~-77  
& 40:1  &1.7
         & \ref{fig:img0023sx} \\
           & 1992.05 & 4.99 &   6    & MERLIN   & 3.6  &168$\times$86 
~-2~~  
& 130:1  & 3.5
         & \ref{fig:img0023cm} \\
\noalign{\smallskip}
0252$-$712 & 1993.14 & 2.29  &   10.25  & 2PHCMGE  &  8.9 & ~32$\times$19 
~-73  & 65:1   &8    
& \ref{fig:img0252s} \\
\noalign{\smallskip}
1151$-$348 & 1991.34 & 2.29  &   11.75  & 35PHC    & 10.7 &~29$\times$13 
~89~   
& 70:1  &4.5
& \ref{fig:img1151sx} \\
           & 1992.24 & 8.42  &   11     & 35PHCM   & 30.3 &~11$\times$5~ 
~81~  
& 35:1  &3.6
& \ref{fig:img1151sx} \\
\noalign{\smallskip}
1306$-$095 & 1993.14 & 2.29  &   10.5   & 32PHCGE  & 10.7 &~48$\times$15 
~-73  
& 20:1  &5.7
& \ref{fig:img1306s} \\
\noalign{\smallskip}
1814$-$637 & 1993.14 & 2.29  &   12     & 32PHCMGE &  9.5 &~32$\times$17 
~-86  
& 50:1  &5.8
& \ref{fig:img1814s} \\
\noalign{\smallskip}
1934$-$638 & \multicolumn{8}{c}{\em 2.29, 4.85 \& 8.42 GHz images in
King, 1994 \& Tzioumis et al. 1996} & \ref{fig:img1934}\\
\noalign{\smallskip}
2135$-$209 & 1991.55 & 2.29  &   12     & 3H       &  3.6  & 
\multicolumn{2}{l}{\em Modelfit only} \\
           & 1992.91 & 4.99  &    6.25  & MERLIN   &  3.6 &163$\times$34 
~10~   
& 130:1   &2.3
& \ref{fig:img2135cm} \\
\hline \hline
\multicolumn{10}{l}{* \footnotesize {Maximum baseline with significant
dectection of each source. No detections to Perth (G) or
Hartebeesthoek (E).}}  
\end{tabular}
\end{center}

\caption{Observations and images presented in this paper.  The dynamic range
         indicates the ratio of the peak brightness in the image to the peak
         in the residual map. The $\sigma_{rms}$ entry refers to the
         residual rms noise in regions devoid of any source emission.
         Entries in the last column refer
         to the figure in which each image appears.\label{tab:obslist}}
\end{table*}

\begin{figure*}
\vspace{-15mm}
\centerline{\psfig{figure=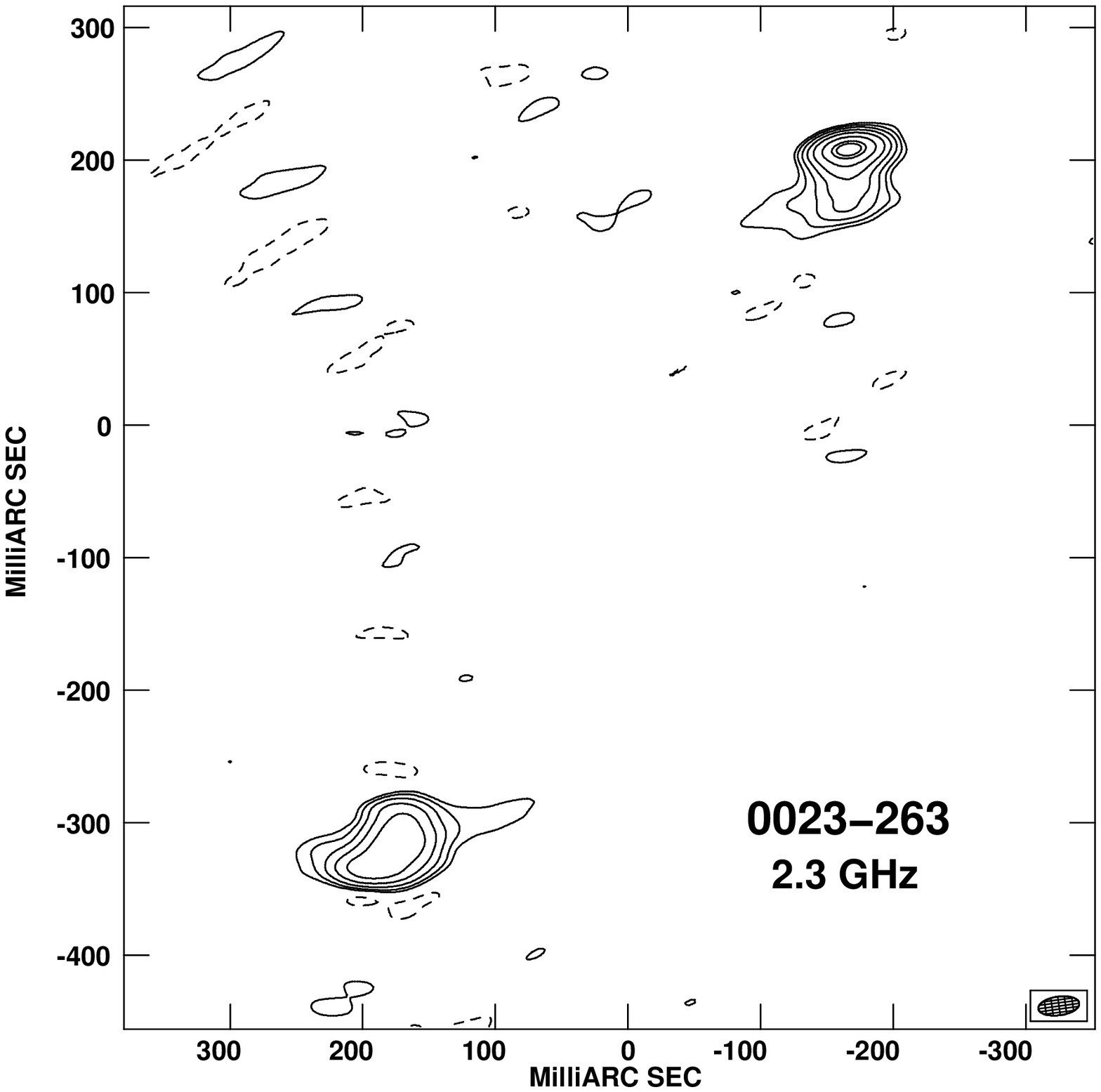,width=0.9\columnwidth,angle=0}
\psfig{figure=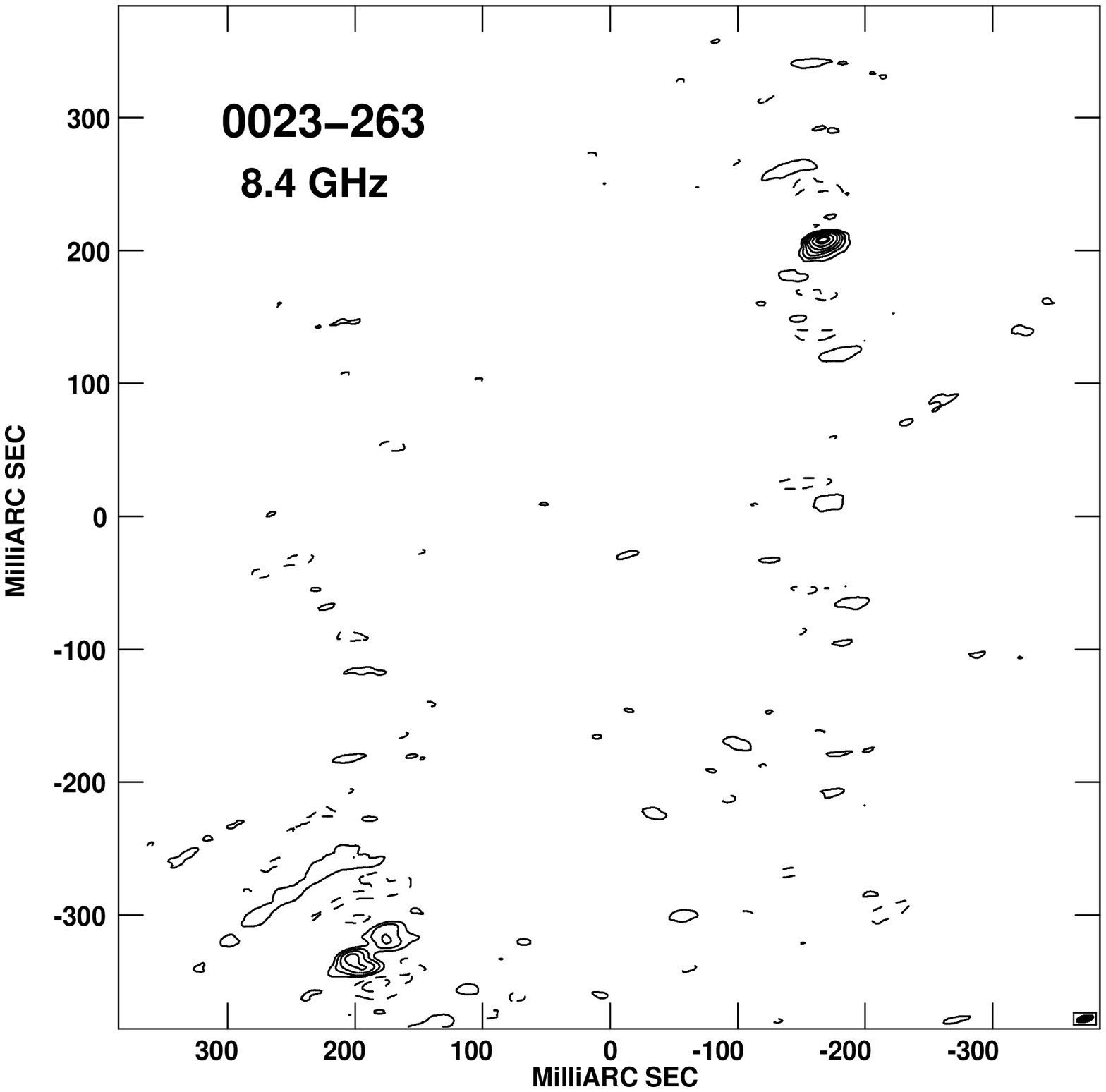,width=0.9\columnwidth,angle=0}}
\vspace{-15mm}
\caption{({\sl Left}) PKS~0023-263 at 2291 MHz from the SHEVE
array. The peak level  is 
1.07 Jy/beam and contours are shown at -2,-1,1,2,4,8,16,35,65,80 \% of
the peak. ({\sl Right)} PKS~0023-263 at 8419 MHz from the SHEVE array. The
peak level is 0.41 mJy/beam and contours are shown at
-2,-1,1,2,4,8,16,35,65,80 \% of the peak.
\label{fig:img0023sx}}
\end{figure*}

\begin{figure}
\vskip 5cm
%\centerline{\psfig{figure=fig4.ps,width=0.9\columnwidth,angle=0}}
\caption{PKS~0023-263 at 4996 MHz from the MERLIN array. The peak level is
1.41 Jy/beam and contours are shown at 
-6,-3,-1.5,-0.75,0.75,1.5,3,6,12,24,48,96 \% of the peak.        
\label{fig:img0023cm}}
\end{figure}

The VLBI observations of all sources were made with the SHEVE network
(Preston et al. 1993; Jauncey et al. 1994) between April 1991 and
February 1993, using the MK2 recording system (Clark, 1973).  The
characteristics of the antennas and receivers in the SHEVE network are
listed in Table~\ref{tab:ants}.  The details of each observation
are given in Table~\ref{tab:obslist}.

All sources were observed at 2.3~GHz, but for $2135-209$ only a single
baseline (Tidbinbilla-Hobart) was obtained in a 12~hour observation.
The sources $0023-263$, $1151-348$ and $1934-638$ were also observed
at 8.4~GHz and $1934-638$ at 4.9~GHz as well. The GPS source
$1934-638$ was observed as part of another program and the details of
the observations and results are presented elsewhere (King 1994;
Tzioumis et al. 1996) while some discussion of the source
properties is presented here. 

The tapes recorded at each station were processed at the Caltech--JPL Block2
correlator in Pasadena and the NRAO AIPS package was used for global fringe
fitting.  The amplitude calibration and editing of the visibilities were
subsequently performed using the Caltech VLBI package. Self-calibration and
imaging of the data were performed using DIFMAP (Shepherd et al. 1994) and the
AIPS packages. The single-baseline data for $2135-209$ was used to obtain a
simple model of the source using the program MODELFIT from the Caltech VLBI
package.

On the very long baselines from the antennas at eastern Australia
(ATNF, DSN, Hobart) to Perth (3000 km) and Hartebeesthoek (9500 km),
only $1934-638$ and VLBI calibrator sources were detected, i.e. all
other sources are completely resolved out at these spacings and
sensitivities. This implies that there are no very compact components
($\theta << $ 3mas) with flux density larger than about 40 mJy, the
detection limit from the intercontinental baselines.

A ``typical'' uv-coverage from the SHEVE array is presented in
Fig.~\ref{fig:uv}, for the source 0252-712 at 2.3 GHz. Note that all 
sources except 1934-638,  were not detected on the long baselines to Perth 
or Hartebeesthoek and uv-tracks are not shown.

The limited number of antennas in the network and the lack of short
baselines limit the dynamic range achievable from these observations.
In particular, for many of the sources the total flux density in the
VLBI images is significantly lower than that measured with lower
resolution observations. This implies that low surface brightness
extended structures with angular sizes between 0.1 and 1 arcsec are
often present but invisible by the VLBI interferometers, as discussed
in the next section on each source. 

For $0023-263$ and $2135-209$ we also present MERLIN observations at
5.0 GHz, which provide information on more extended structures, not
detected by the VLBI observations.

The new images of the six sources observed in this program are
presented in Figs.~\ref{fig:img0023sx} through \ref{fig:img2135cm}.
Restoring beams together with  the noise measure on the image plane
are given in Table~\ref{tab:obslist}. 

The overall size of most sources in this sample was not previously
well determined at radio wavelengths.  The largest angular/linear size
for each source was estimated from the images, adopting a ``low
resolution'' approach (cf. Dallacasa et al. 1995). Most of the sources
show distinct and well separated ``lobes'', and extended structure can
be seen in most of them. Their sizes and flux densities were estimated
from the images and are summarised in Table~\ref{tab:components}. The
component angular dimensions were generally estimated directly from
the images (except were otherwise stated) by considering the lowest
reliable contour; they are given as full-width axis sizes. This
approach was adopted because almost all components show extended
structures and it is then difficult to fit single Gaussian
components. 
The measurements provided values well in excess of the beam size, and
they were not deconvolve given they would not change significantly;
therefore, all sizes in Table~\ref{tab:components} should be treated
as upper limits.  

The separation between the strongest component in each of the lobes
was also measured from the images.  
This separation is a very sensitive parameter since it is well determined 
by the ``beating'' in the visibilities and can be used to search for any
source expansion in time, even with sparse uv coverage.  
The results are presented in Table~\ref{tab:size}. 

For the two sources observed at 2.3 and 8.4 GHz, a two-point spectral
index for each lobe has also been determined.

\begin{table}
\begin{center}
\begin{tabular}{lccccc}
\hline\hline
      &    &    & \multicolumn{3}{c}{Component separation} \\
Name  & $\theta_{max}$ &  Lin. size & $\theta$ & $l$ & Orientation \\
      & mas  & pc & mas  &  pc    &  $^{\circ}$  \\
\hline  
           &   &      &       &     &  \\
$0023-263$ & 680 & 1970 & 654 & 1900 & -34 \\
$0252-712$ & 240 &  900 & 145 &  540 &   7 \\
$1151-348$ & 170 &  425 &  91 &  228 &  72 \\
$1306-095$ & 460 & 1600 & 373 & 1290 & -41 \\
$1814-637$ & 410 &  328 & 239 &  191 & -20 \\
$1934-638^*$ &  70 &  140 &  42 &   84 &  89 \\
$2135-209$ & $\sim$250 & $\sim$1000  & 167 & 650 &  52 \\
\hline
\multicolumn{6}{l}{\small $^*$ Tzioumis et al. 1996}
\end{tabular}
\end{center}
\caption{Overall angular and linear sizes of the sample sources and
separation of strongest components in the two lobes.\label{tab:size}} 
\end{table}

\begin{table*}
\begin{center}
\begin{minipage}{\textwidth}
\begin{tabular}{ll c rrrrrrc}
\hline\hline
Source & Comp. & $\nu$ & $\theta_1$ & $\theta_2$ &  Lsize1 & Lsize2 & 
$S_{peak}$ & $S_{tot.}$ & $\alpha^{8.4}_{2.3}$ \\
       &   & GHz   & mas  & mas & pc  &  pc    &  Jy/beam  & Jy & \\
\hline  
$0023-263$ & NW & 2.3 & 80 & 75 & 230 & 220 & 1.07 & 1.6 & -0.7 \\
           &    & 8.4 & 40 & 23 & 110 & 65  & 0.41 & 0.65 &  \\
           & SE & 2.3 & 100 & 90 & 290 & 260 & 0.33 & 1.3 & -1.3 \\
           &    & 8.4 & 60 & 30 & 175 & 85 & 0.047 & 0.25 & \\
\noalign{\smallskip}
$0252-712$ & North & 2.3 & 110 & 90 & 400 & 330 & 1.14 & 2.1 &  \\
           & South &     & 120 & 90 & 440 & 330 & 0.47 & 1.4 &  \\
\noalign{\smallskip}
$1151-348$ & NE & 2.3 & 85 & 45 & 210 & 110 & 1.57 & 2.1  & -0.6  \\
           &    & 8.4 & 40 & 30 & 100 & 75 & 0.437 & 0.94 &  \\
           & SW & 2.3 & 85 & 45 & 210 & 110 & 0.78  & 1.24  & -1.0 \\
           &    & 8.4 & 35 & 20 & 90 & 50 & 0.12  & 0.36 & \\
\noalign{\smallskip}
$1306-095$ & SE & 2.3 & 110 & 70 & 380 & 240 & 0.304 & 0.76 &  \\
           & NW &     &  80 & 50 & 275 & 170 & 0.107 & 0.15 &  \\
\noalign{\smallskip}
$1814-637$ & North & 2.3 & 90 & 80 & 70  & 65 & 1.58 & 2.8 &  \\
           & South &  & 130 & 70 & 100 & 55 & 0.380 & 1.8 &  \\
\noalign{\smallskip}
$1934-638$\footnote{Components and spectra for $1934-638$
are given in Tzioumis et al. 1996.} \\
$2135-209$\footnote{Component sizes from fitting
Gaussians (FWHM)  since the beam is comparable to the source size and
 components cannot be estimated graphically.}
           & NE & 5.0 & 25 & 9 & 100 & 350 & 1.02 & 1.1 &  \\
           & SW &  & 90 & 40 & 350 & 155 & 0.2 & 0.36 &  \\
\hline
\end{tabular}
\end{minipage}
\end{center}
\caption{Size and flux density of the components in the
images. Component  angular sizes are estimated from the images
(full width), except were otherwise noted. \label{tab:components}} 
\end{table*}

\section{Comments on individual sources}

In the following, a description of the morphology of each source is given:

\paragraph{0023-263} This radio source has been identified with a galaxy of
$m_v=19.5$ by WP85 at a redshift of 0.322 (Tadhunter et al. 1993; di
Serego Alighieri et al. 1994).  The VLBI images at 2.3 and 8.4
GHz are shown in Fig.~\ref{fig:img0023sx}.

At both VLBI frequencies this radio galaxy shows a basic double
structure with a separation of $\sim 650$ mas ($\sim$1.9kpc) in
p.a. $-34^\circ$ between the peak brightness of each lobe. Both radio
emitting regions are complex and extended structures are evident. In
particular, the south-eastern component is resolved into two
sub-components embedded into more diffuse emission. Thus, it is not
possible to fit simple Gaussian components, and sizes for each lobe
are estimated from the images. Both lobes are less than $\sim$100 mas
($\sim$ 0.3kpc) in extent and the south-eastern one has an overall
spectral index ($\alpha$=--1.3) steeper than the north-western
component ($\alpha$=--0.7), possibly also because it is more resolved.

The flux density detected in the VLBI images accounts for only about 40\% of
the total source flux density at each frequency.  This indicates that there is
undetected extended scale structures in the range 0.1 to 1 arcseconds.  The
same fraction of flux density is undetected at both frequencies, suggesting
that the underlying diffuse structure has similar spectral index ($\sim
- 0.9$) to the components visible in the images.  No flat spectrum component
indicative of a core is detected between the lobes at the detection limits
(5$\sigma$) of $\sim$20~mJy/beam at 2.3~GHz and $\sim$10~mJy/beam at 8.4 GHz. 
We decided to use 5 $\sigma$ as detection limit for the cores (in this
and in the following sources), given that, the complexity of the radio
emission in relation to the limited uv-sampling, increases the incidence of
clean artifacts at lower levels.

The MERLIN image at 5 GHz (Fig.~\ref{fig:img0023cm}) with much lower
resolution shows the basic double-lobe structure of the source and the
extended emission surrounding the lobes. About 90\% of he total flux density 
is detected (3.4 Jy out of 3.8 Jy) indicating little extended emission at 
arcsecond scales.

\begin{figure}
\vspace{-15mm}
\centerline{\psfig{figure=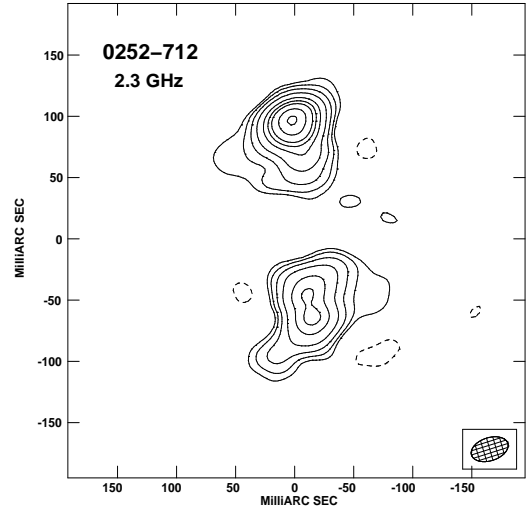,width=0.9\columnwidth,angle=0}}
\vspace{-15mm}
\caption{PKS~0252-712 at 2291 MHz from the SHEVE array. The peak level is 
1.14 Jy/beam and contours are shown at
-1.5,1.5,3,6,12,24,36,48,72,96 \% of the peak.     
\label{fig:img0252s}}
\end{figure}

\paragraph{0252-712} This object has been identified with a galaxy of
$m_V$=20.9 and redshift $z=0.566$ (di Serego Alighieri et al. 1994). 
The VLBI image at 2.3 GHz (Fig.~\ref{fig:img0252s}) shows a
North-South double structure quite symmetric; the peak regions are 
separated by $\sim$145~mas ($\sim$0.54~kpc).  Both components appear
to be resolved  and account for more than 95\% of the whole flux
density of 3.7~Jy at this frequency. No ``core'' component is
detected between the lobes (5$\sigma$=40~mJy) and no spectral
information is available on the detected components.

\begin{figure*}
\vspace{-15mm}
\centerline{\psfig{figure=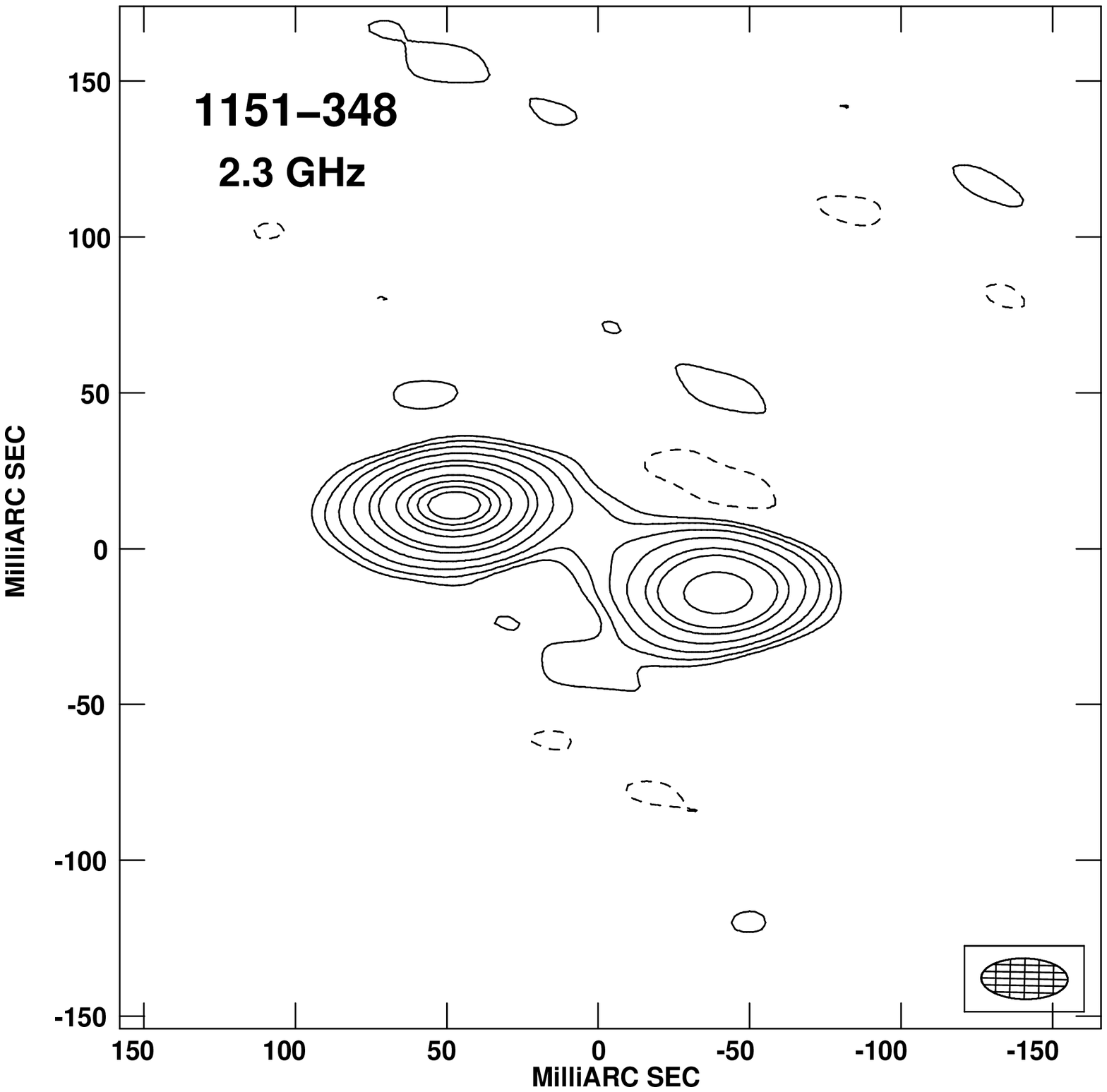,width=0.9\columnwidth,angle=0}\psfig{figure=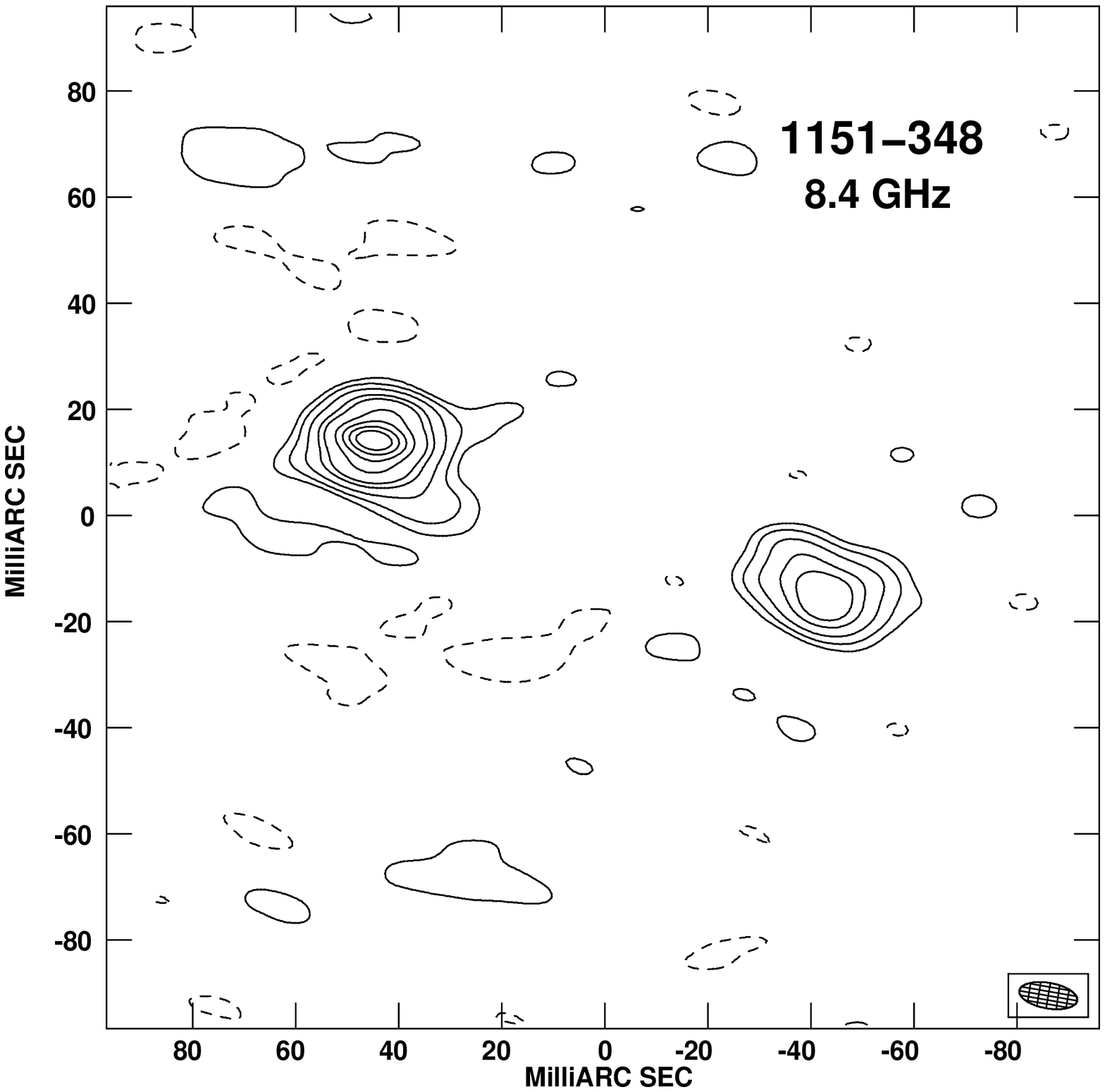,width=0.9\columnwidth,angle=0}}
\vspace{-15mm}
\caption{({\sl Left}) PKS~1151-348 at 2291 MHz from the SHEVE array. The peak level is
1.57 Jy/beam and contours are shown at
-0.75,0.75,1.5,3,6,12,18,35,50,65,80 \% of the peak. ({\sl Right}) PKS~1151-348 at 8421 MHz from the SHEVE array. The peak level is
0.44 Jy/beam and contours are shown at
-1.5,1.5,3,6,12,18,35,50,65,80 \% of the peak.
\label{fig:img1151sx}}
\end{figure*}

\paragraph{1151-348} This object has been identified as a quasar with
$m_v=17.8$ from WP85 and $z=0.258$ (Jauncey et al. 1978).  Both 2.3
and 8.4 GHz images are available
(Fig~\ref{fig:img1151sx}), showing two components
with separation of 91 mas ($\sim$ 0.23 kpc) and p.a.$=72^\circ$ and
possibly some extended emission between them. The peak separation and
orientation are identical (to within 0.5 mas) at both frequencies,
unambiguously registering the two components. 

Both lobes are less than $\sim$85 mas ($\sim$ 0.2kpc) in extent and
the south-western one has an overall spectral index ($\alpha$=--1.0)
steeper than the north-eastern component ($\alpha$=--0.6). About 70\%
of the total flux density is detected in the VLBI components and this
fraction is the same at both frequencies. Thus the underlying
undetected flux density has a spectral index ($\sim$--0.7) similar to
the two aforementioned components, possibly indicating that this emission
is simply resolved, diffuse extension of these lobes.
 
No flat spectrum component, possibly indicative of a core, is detected
between the lobes, at the detection limit (5$\sigma$) of
$\sim$20~mJy/beam.

\paragraph{1306-095} It is identified with a galaxy of magnitude
$m_V$=20.5 and redshift $z$=0.464 (di Serego Alighieri et al. 1994).

\begin{figure}
\vspace{-15mm}
\centerline{\psfig{figure=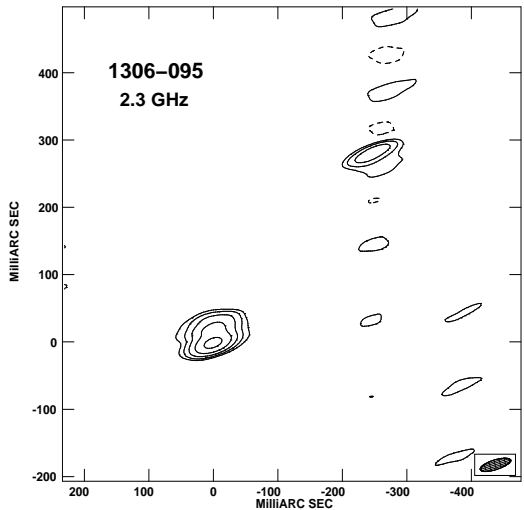,width=0.9\columnwidth,angle=0}}
\vspace{-15mm}
\caption{PKS~1306-095 at 2291 MHz from the SHEVE array. The peak level is
0.30 Jy/beam and contours are shown at
-5,5,10,20,40,80 \% of the peak.    
\label{fig:img1306s}}
\end{figure}

This source is at low absolute declination and the uv coverage is rather poor,
which implies high sidelobe levels ($\sim$50\%)
and a low dynamic range image. The radio emission is
dominated by two distinct components separated by about 373 mas
($\sim$1.3 kpc) in p.a. $\sim -40^\circ$
(Fig.~\ref{fig:img1306s}). The southern component is much stronger,
accounting for more than 80\% of the detected flux density.  Only
about 30\% of the total flux density is detected by the VLBI
components, indicating the presence of much diffuse emission
completely resolved out by the VLBI baselines. It is likely that in
Fig~\ref{fig:img1306s} only the hot-spots of two mini-lobes are
visible.

\begin{figure}
\vspace{-15mm}
\centerline{\psfig{figure=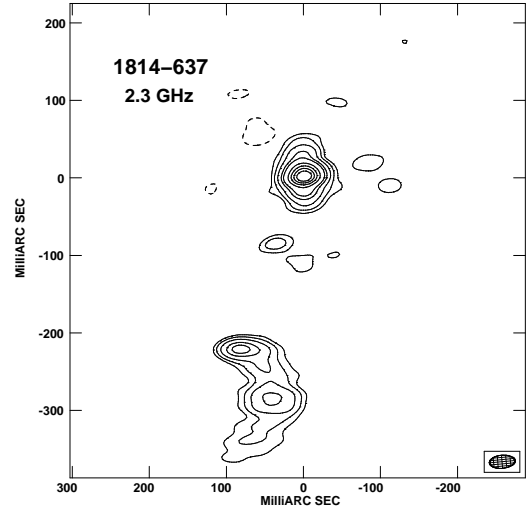,width=0.9\columnwidth,angle=0}}
\vspace{-15mm}
\caption{PKS~1814-637 at 2291 MHz from the SHEVE array. The peak level is
1.7 Jy/beam and contours are shown at
-1.5,1.5,3,6,12,18,35,50,65,80 \% of the peak.
\label{fig:img1814s}}
\end{figure}

\paragraph{1814-637} 
It is identified with a galaxy of $m_V=18.0$ with $z=0.063$ by WP85.
The VLBI image at 2.3 GHZ shows the most complex structure of all
sources in this sample (Fig.~\ref{fig:img1814s}).  The source still
shows a basic double-lobed structure aligned almost
North-South. However, the southern region is dominated by two
components with similar brightness and much extended emission.  The
northern region shows a prominent bright component and extended
emission, with possible North-South symmetrical extensions, but the
overall extent of this component is still less than $\sim$90 mas
($\sim$70 pc). \\
Just over 50\% of the total flux density of the source is detected,
indicating the presence of further, much more diffuse extended
component. The overall source size is still less than about 0.5
arcsecond and no arcsecond-scale components are detected with ATCA
observations. \\ 
There is also a weak component between the two major lobes at 
the 5\% level, which corresponds to a 15$\sigma$ detection. This
can be just a peak in the underlying diffuse component or could
indicate the presence of a core between the two lobes. VLBI
observations at another frequency are needed to determine the spectrum
of this component and decide whether it is the core.

\begin{figure*}[t]
\centerline{\psfig{figure=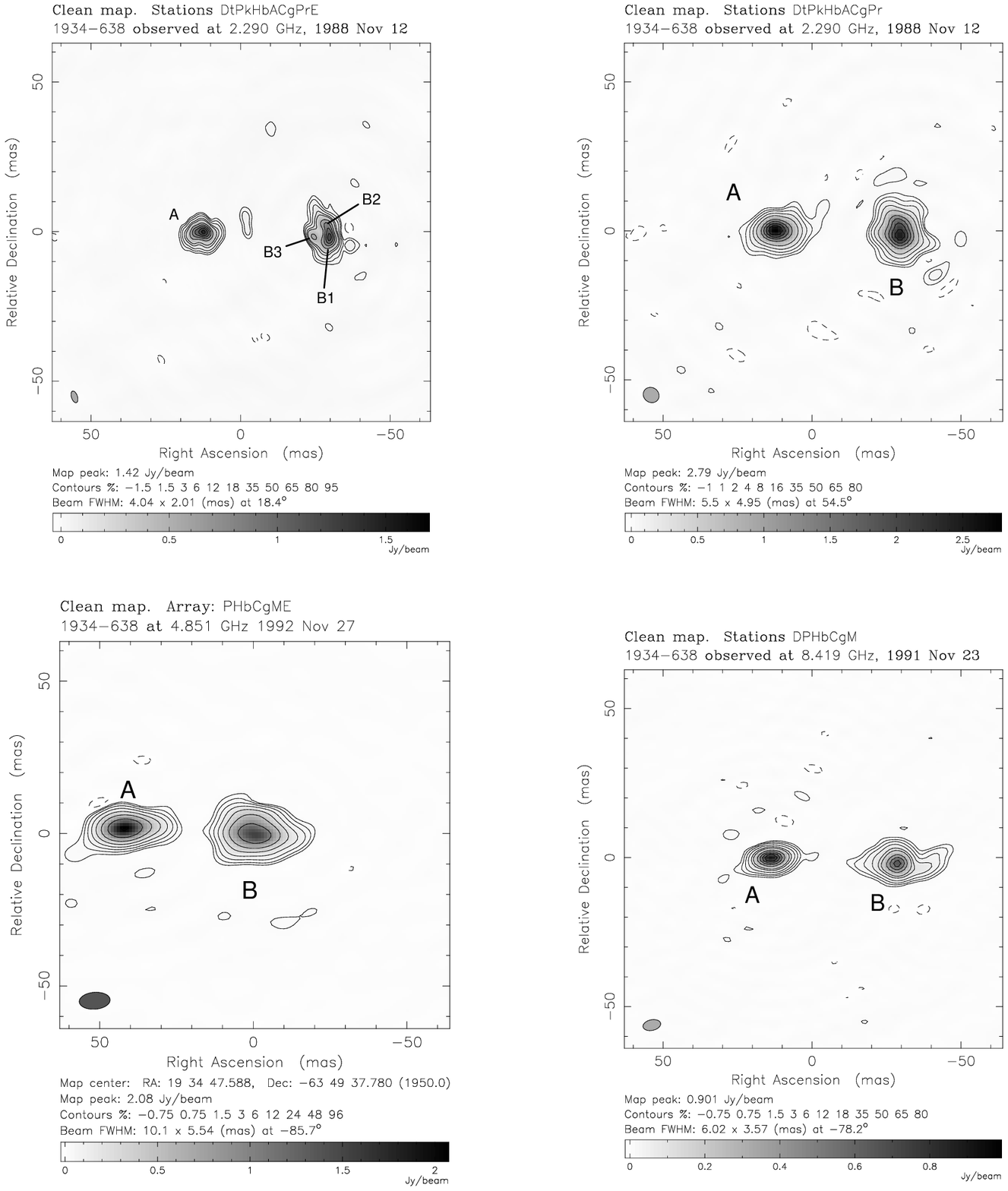,width=\textwidth,angle=0}}
\caption{VLBI images of PKS~1934-638 at 2.3, 4.8 and 8.4 GHz (from Tzioumis 
et al. 1997). The top-right 
image is at a higher resolution because it includes Hartebeesthoek data. 
The other three images have approximately the same resolution of about 5 mas.
         \label{fig:img1934}}
\end{figure*}

\paragraph{1934-63} 
This compact radio source has been identified as a galaxy with
$m_V=18.4$ by Kellerman (1966) and has redshift $z=0.183$ (Penston \&
Fosbury 1978).  A VLBI image at 2.3~GHz from observations carried out
in 1982 has been presented in Tzioumis et al. (1989) and preliminary
images at 2.3, 4.8 and 8.4 GHz have also been presented elsewhere
(King 1994; Tzioumis et al. 1996). The knowledge about the pc-scale
structure of this source is summarised in Fig.~\ref{fig:img1934}.
The source appears as a very compact double, with component separation
of 42 mas ($\sim$ 0.084 kpc) in p.a. 89$^\circ$, and this separation
has not changed for over 15 years.  As reported in Tzioumis et
al. (1996), this yield to an upper limit to the expansion velocity of
$0.2c$.  The total flux density of the source is
detected by the VLBI components and there is no definite detection of
a flat-spectrum core component at a level of a few \% of the peak at
any frequency.

\paragraph{2135-209}
This source is a galaxy with $m_v=19.4$ and  $z=0.635$ (WP85 and di Serego
Alighieri et al. 1994) .

\begin{figure}
\vspace{-15mm}
\centerline{\psfig{figure=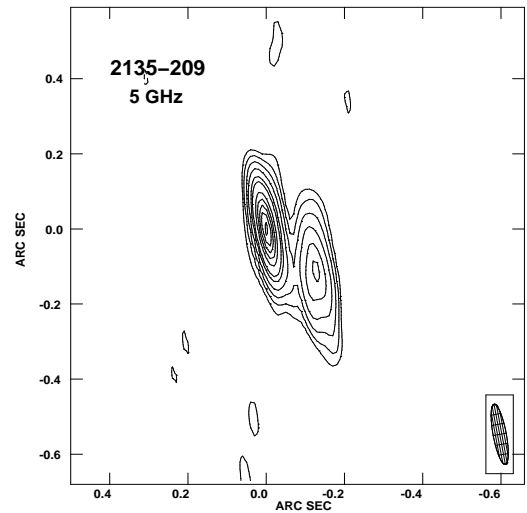,width=0.9\columnwidth,angle=0}}
\vspace{-15mm}
\caption{PKS~2135-209 at 4993 MHz from the MERLIN array. The peak level is
1.02 Jy/beam and contours are shown at
-0.75,0.75,1.5,3,6,12,18,35,50,65,80,95 \% of the peak.
\label{fig:img2135cm}}
\end{figure}

No VLBI image exists for this source and here we present a MERLIN 5 GHz image
(Fig.~\ref{fig:img2135cm}). The source shows a double lobed very asymmetric
structure; the resolution of MERLIN at this southern declination is poor 
($163 \times 34$ mas in the case of 2135-209) to allow further consideration
on the morphology. Because of the limited resolution compared to the source
extension, the component parameters were derived by means of Gaussian
fitting. The North--East dominant component is marginally resolved and
accounts for more than 80\% of the detected flux density. The total flux
density of about 1.5~Jy was detected in the MERLIN image, indicating no extra
extended components exist.

From one-baseline VLBI observations at 2.3 GHz it is possible to fit
models to the double structure. The strong north-east component is
unresolved while the south-west component is extended. The separation
between the peaks in the two lobes is well determined at 167 mas
($\sim$0.65 kpc), at a position angle of 52$^\circ$, close to the
position angle and separation determined from the MERLIN image. About
85\% of the total flux density is detected on this VLBI baseline.

\section{Discussion}

The main aim of these high resolution observations was the
morphological classification of a group of CSS sources found in a
larger, flux-limited sample mainly consisting of extended radio
sources. Like the majority of CSS sources studied in previous works,
the objects observed here show mainly a double-lobed structure, often
asymmetric. 
The structures revealed by these observations are generally resolved
and they may well represent the lobes of small-sized powerful radio
sources. In fact, for the two sources observed at both 8.4 and 2.3
GHz, only steep spectrum components have been detected. The present
observations do not reveal any definite core candidates, but in one
case ($1814-637$) there is a weak third component between the 
lobes which could be suggested as a possible core. Weak tails
are found pointing towards the source centre in a few cases,
further suggesting that we are seeing two-sided emission from lobes
and possibly hot-spots.

Generally, these images do not show  many highly distorted or complex
structures, which could be ascribed either to the
interaction between the radio emitting plasma and the ambient medium
or to projection effects.  Even in the object which shows the most complex
structure  ($1814-637$),  the basic double-lobed morphology is still
clearly evident.  Source linear sizes are in the range 0.08--2.9~kpc,
and hence all sources in this sample can be classified as compact
symmetric (CSO) or medium-sized symmetric (MSO) objects  if we
assume that an undetected core sits between the lobes revealed by the
present observations.

Dallacasa et al. (1995) have investigated a sample of northern CSS
sources (using MERLIN and EVN) selected from the Peacock \& Wall
catalogue (1981). As their sample and the present sample have the same
selection and observing frequencies, it is expected that sources in
both samples should have, on average, similar characteristics.
CSS sources selected from the Wall \& Peacock sample tend to have a
higher turnover frequency and smaller size (e.g. Dallacasa et al.
1995) because of the higher selection frequency (\ie\ 2.7GHz) compared
with the 3CR sample (178 MHz).

The size of the sources in the present sample range between $\sim$0.08 and
$\sim$2.9~kpc, similar to the range found in the sample studied in Dallacasa
\etal (1995), and smaller than the 3CR samples (Fanti \etal
1990 and references therein). Turnover frequencies are in the range
around 0.1 -- 0.2 GHz or below, with the exception of 1934-63 that
peaks at $\sim 1.4$GHz. As expected from the correlation between the
linear size of the source and the turnover frequency (Fanti et al.
1990), the smallest source  in our sample is the GPS source
$1934-638$, that has the turnover at higher frequency.

Dallacasa et al. (1995) also find that most of galaxies their PW CSS
sample show weak core candidates at \ltae 2\% of the total
flux density, in contrast with the 3CR samples, where, however, the
dynamic range was smaller and comparable to these achieved for most of
the sources presented in this paper. 

Of the seven CSS/GPS objects presented in this paper, four have been
searched for HI absorption (0023-26, 1814-63, 1934-63 and 2135-20).
The results have been summarised in Morganti et al. (2001) while
the preliminary results from new VLBI observations for 1814-63 have
been presented in Morganti et al.(2000).  

Of the four objects, three are detected.  The only one undetected is
2135-209, but  this result is not too surprising given that it is
classified as broad-line   galaxy. It is interesting, however, that
the most impressive HI absorption is   detected in the object with   the
most complex radio structure in our small   sample, 1814-63.   The
deep HI absorption appears extended and cover most of the
source with the exception of the southern part of the southern   lobe.
The peak optical depth goes from $\sim 30$\% in the southern lobe to
$\sim$ 10\% in the northern region.  The component with lower optical
depth   (ranging between 2 and 4\%) is observed {\sl only} against the
northern   region. Using the available systemic velocity ($V_{\rm
hel}=19350$ \kms,   Morganti et al. 2001), most of the \HI\ absorption
appears to be   blueshifted. This can be explained by the presence of
extended gas, possibly   surrounding the lobes, interacting/expanding
with the radio plasma.   Therefore, in the case of 1814-63, part of
the radio morphology could be   affected by a strong interaction with
the ISM. 

Overall, the properties of the sources discussed here are
consistent with the results from the analysis of other samples of CSS
sources.
In a forthcoming paper we will discuss their properties in wavebands
different from radio, compare them with those of extended radio
sources and verify whether they support  and
{\sl  ``youth''} scenario.

{}

%\newpage

\begin{thebibliography}{}
\bibitem{} Axon D.J., Capetti A., Fanti R. et al. 2000, AJ 120, 2284
\bibitem{} Bicknell J., Dopita M., O'Dea C. 1997, ApJ 485, 112
\bibitem{} Clark, B.G., 1973, Proc. IEEE, 61, 1242 
\bibitem{} Conway J.E.  1996, in " The second Workshop on Gigahertz
Peaked Spectrum and Compact Steep Spectrum Radio Sources", Eds.
Snellen, Schilizzi et al.  M.N.  Publ JIVE, Leiden p.198
\bibitem{} Dallacasa D., Fanti C., Schilizzi R.T., Spencer R.E., 1995, A\&A
  295, 27 
\bibitem{} Fanti C., Pozzi F. Fanti R. et al. 2000, A\&A 358, 499
\bibitem{} Fanti R., Fanti C., Schilizzi R.T. et al. 1990,  A\&A 231, 333
\bibitem{} Fanti, C., Fanti, R., Dallacasa D. et al.  1995 A\&A 302, 317 
\bibitem{} Fanti C., 2000, "Proceedings of the 5th European VLBI Network
  Symposium" J.E. Conway, A.G. Polatidis, R.S. Booth and Y.M. Pihlstr\"{o}m
  ISBN 91-631-0548-9, p. 73
\bibitem{} Gelderman R. \& Whittle M. 1994, ApJS 91,491
\bibitem{} Jauncey D.L., Wright A.E., Peterson B.A. \& Condon J. 1978, ApJ 219, L1 
\bibitem{} Jauncey D.L. \& the SHEVE Team 1994, in "Very high angular resolution
imaging", IAU158, J. G. Robertson and William J. Tango., eds., Dordrecht:
Kluwer, p. 131
\bibitem{} Kellerman K.I.  1966 ApJ 146, 621 
\bibitem{} King E. 1994 PhD Thesis Univ. of Tasmania
\bibitem{} Morganti, R., Killeen, N.E.B. \& Tadhunter, C.N. 1993, MNRAS, 263,
1023
\bibitem{} Morganti R., Oosterloo T., Tadhunter C.N., van Moorsel, Killeen N. \&
  Wills   K.A.  2001, MNRAS 323, 331
\bibitem{} Morganti R.,  Oosterloo T.A., Reynolds J., Tadhunter C.N. \& Migenes
V.  1997,  MNRAS, 284, 541
\bibitem{} Morganti R., Oosterloo T.A., Tadhunter C.N. et al.  1999, A\&A Suppl. 140, 355
\bibitem{} Morganti R., Oosterloo T., Tadhunter C.N. et al.  2000, 'Proceedings of the 5th European VLBI Network Symposium'
  J.E. Conway, A.G. Polatidis, R.S. Booth and Y.M. Pihlstr\"{o}m ISBN
  91-631-0548-9, p. 111 (astro-ph/0010482)
\bibitem{} Murgia M. et al. 1999 A\&A 345, 769
\bibitem{} O'Dea, C.P., Baum, S.A. \& Stanghellini, C., 1991, ApJ 380, 66 
\bibitem{} O'Dea, C.P. 1998 PASP 110, 493
\bibitem{} O'Dea C.P., de Vries W.H., Koekemoer A.M. et al. 2002 AJ accepted 
\bibitem{} Owsianik I. \& Conway J.E.1998, A\&A 337, 69
\bibitem{} Padovani P., Morganti R., Siebert J., Cimatti A. \& Vagnetti F.
  1999, MNRAS, 304, 829
\bibitem{} Peck A.B., Taylor G.B. 2001, ApJ 554, L147
\bibitem{} Penston M.V. \& Fosbury R.A.E. 1978, MNRAS 183, 479 
\bibitem{} Pihlstr\"om Y.M. 2001, PhD. Thesis, Chalmers University \& Onsala
Observatory
\bibitem{} Preston \& the SHEVE Team, 1993, in {\it Sub-arcsecond Radio
Astronomy}, eds. R.J. Davis and R. S. Booth, Cambridge University Press, 1993., p.428 
\bibitem{} di Serego Alighieri S., Danziger J.I., Morganti R. \& Tadhunter C.,
1994, MNRAS 269, 998 
\bibitem{} de Vries W., O'Dea C.P., Baum S.A. et al. 1998, ApJ 503, 138
\bibitem{} Sanghera H.S. 1993, PhD Thesis, Univ. of Manchester
\bibitem{} Shepherd, M. C., Pearson, T. J., \& Taylor, G. B. 1994, BAAS, 26, 987 
\bibitem{} Siebert, J., Brinkmann W., Morganti, R. et al.  1996, MNRAS 279, 1331 
\bibitem{} Silk J., Rees M.J. 1998, A\&A 331, L1
\bibitem{} Spencer, R.E., Schilizzi, R.T., Fanti et al. 1991, MNRAS,
250, 225 
\bibitem{} Tadhunter, C.N., Morganti, R., di Serego Alighieri, S., Fosbury,
R.A.E. \& Danziger, I.J.  1993, MNRAS, 263, 999
\bibitem{} Tadhunter C.N., Shaw M. \& Morganti R. 1994. MNRAS, 271, 807 
\bibitem{} Tadhunter C.N., Morganti R., Robinson A. et al. 1998,  MNRAS, 298, 1035
\bibitem{} Trussoni E., Vagnetti F., Massaglia S. et al.  1999, A\&A 348, 437
\bibitem{} Tzioumis et al. 1996, in {\it The second Workshop on Gigahertz
Peaked Spectrum and Compact Steep Spectrum Radio Sources}, Eds.
Snellen, Schilizzi et al.  M.N.  Publ JIVE, Leiden p. 58
\bibitem{} Venturi T., Morganti R., Tzioumis A. \& Reynolds J., 2000 A\&A 363, 84
\bibitem{} Wall J. \& Peacock J. 1985, MNRAS 216, 173 

\end{thebibliography}
\end{document}